\input phyzzx\def\e{\adveq\eqno{\rm (\chapterlabel\the\equanumber)}}

\def\adveq{\global\advance\equanumber by 1}
\def\myeq{{\rm \chapterlabel\the\equanumber}}

\def\manyeq#1{\eqalign{#1}\e}

\def\semidirect{\mathrel{\raise0.04cm\hbox{${\scriptscriptstyle |\!}$
\hskip-0.175cm}\times}}


\def\ref#1{$^{[#1]}$}

\def\r#1{$[\rm#1]$}

\def\e{\adveq\eqno{\rm (\chapterlabel\the\equanumber)}}

\def\adveq{\global\advance\equanumber by 1}
\def\myeq{{\rm \chapterlabel\the\equanumber}}

\def\manyeq#1{\eqalign{#1}\e}

\def\semidirect{\mathrel{\raise0.04cm\hbox{${\scriptscriptstyle |\!}$
\hskip-0.175cm}\times}}


\def\ref#1{$^{[#1]}$}

\def\r#1{$[\rm#1]$}

%

%

\date{February, 2007}
\titlepage
\title{On the Algebraic Approach to Solvable Lattice Models.}
\author{Andrei Babichenko${}^{(1,2)}$ and Doron Gepner${}^{(1)}$}
\line{}
\line{\hfill 1) Department of Particle Physics\hfill}
\line{\hfill Weizmann Institute\hfill}
\line{\hfill Rehovot 76100, Israel\hfill}
\line{}
\line{\hfill 2) Racah Institute of Physics\hfill}
\line{\hfill The Hebrew University\hfill}
\line{\hfill Jerusalem 91904, Israel\hfill}
\abstract
We develop an algebraic approach to solvable lattice models based on a
chain of algebras obeyed by the models. In each subalgebra we use a unit,
giving a chain of ideals. Thus, we divide the models into distinct sectors which do
not mix. This method gives the usual Bethe ansatz results in cases it is known, but
generalizes it to non integrable models. We exemplify the method on the 
Temperley--Lieb and Fuss--Catalan algebras. For the Fuss--Catalan algebra we show that the
ground state energy is zero and there is a mass gap of one for $\alpha>\sqrt2$,
and that for $\alpha=1$ we seem to get an RCFT as the scaling limit.
\endpage
\par

\section{Introduction.}
Solvable lattice models in two dimensions have attracted considerable
attention (for a review see \REF\Baxbook{Baxter R., ``Solvable lattice 
models", Book (1982).}\r\Baxbook).
This due to a large extent owing to their solvability along with
the fascinating interplay with quantum field theory close to the
critical points. Here we use the algebraic structure of such models
to study their spectrum. Our main results are about the Fuss--Catalan
models where we find a mass gap of $1$ and a ground state energy zero 
in the principal regime.
For $\alpha=1$ we conjecture to get a series of conformal field 
theories and no mass gap.

\section{The algebras.}
In the process of solving various lattice models one encounters
different algebras. It is an old idea of Baxter 
\REF\Baxt{Baxter R. J., J. Stat. Phys. 28 (1982) 1.}
ref. \r\Baxt\ that the 
algebraic
structure itself can be used to solve the models completely.
This is the line of thought that we will follow here.
We develop further the interesting ideas of Levy on the subject
\REF\Levy{Levy D., Phys. Rev. Lett. 64 (1990) 499.}\r\Levy.

The best known example is the Temperley--Lieb algebra 
\REF\Temper{Temperley H and Lieb E H, Proc. R. Soc. A 532 (1971) 609.}\r\Temper.
It is defined on the $n$ points of the model by,
$$\manyeq{e_i^2&=\alpha e_i\cr
  e_i e_{i\pm 1} e_i&=e_i\cr
  e_i e_j&=e_j e_i\quad{\rm for\ } |i-j|\geq2,\cr}$$
where the $e_i$ are the generators of the algebra,
and $i=1,2,\ldots,n$.
Originally, this algebra was
used to solve thge $q$ states Potts model where $\alpha=q^{1/2}$.
Another interesting model is the Heisenberg spin chain where one
takes, 
$${\cal H}=\sum_{i=1}^n e_i.\e$$

Another algebra of this kind is the Fuss--Catalan algebra (FC),
\REF\Bisch{Bisch D and Jones V., Inv. Math. 128 (1997) 89.}
\REF\Di{P. Di Francesco, Nucl. Phys. B 532 (1998) 609.}
\r{\Bisch,\Di}, 
which is a generalization of the Temperley Lieb algebra (TL) to
more than one color. The generators here are $e_i^{(a)}$ where 
$a$ labels the colors, $a=1,2,\ldots,f$ where $f$ is the
number of colors. $f=1$ is just the usual Temperley--Lieb algebra.

The two colors Fuss--Catalan algebra is defined by,
$$\manyeq{(e_i^{(1)})^2&=\alpha e_i^{(1)}\cr
(e_i^{(2)})^2&=\alpha^2 e_i^{(2)}\cr
e_i^{(1)} e_i^{(2)}&=\alpha e_i^{(2)}\cr
e_i^{(1)} e_{i+1}^{(1)}&=e_{i+1}^{(1)} e_i^{(1)}\cr
e_i^{(1)} e_{i\pm1}^{(2)} e_i^1&=e_i^{(1)} e_{i\pm1}^{(1)}\cr
e_i^{(2)} e_{i\pm1}^{(1)} e_i^{(2)}&=\alpha e_i^{(2)}\cr
e_i^{(2)} e_{i\pm1}^{(2)} e_i^{(2)}&=e_i^{(2)}\cr
e_i^{(1)} e_{i\pm1}^{(2)} e_i^{(2)}&=e_{i\pm1}^{(1)} e_i^{(2)}\cr
e_i^{(2)} e_{i\pm1}^{(2)} e_i^{(1)}&=e_i^{(2)} e_{i\pm1}^{(1)}.\cr}$$
Here $\alpha$ is a free parameter defining the algebra, and we assume
that $i=1,2,\ldots, n$ is the number of points. 
We take the Hamiltonian to be,
$${\cal H}=\sum_{i=1}^n {e_i^{(1)} \over \alpha}+
{e_i^{(2)}\over \alpha^2-2}.\e$$
We take this Hamiltonian so as to become an integrable model,
where the transfer matrix obeys the Yang--Baxter relation, ref. \r\Di.

For more than two colors the algebra is given by
the relations,
$$\manyeq{e_i^{(m)} e_i^{(p)}&=\alpha^{\min(m,p)} e_i^{(\max(m,p))}\cr
  e_i^{(m)} e_j^{(p)}&=e_j^{(p)} e_i^{(m)},\qquad {\rm if\ }|i-j|>1 {\rm\ or\ }
  j=i\pm1{\rm\ and\ } m+p\leq f.\cr
  e_i^{(m)} e_{i\pm1}^{(p)} e_i^{(m)}&=\alpha^{(f-p)} e_i^{(m)} e_{i\pm1}^{(f-m)}\qquad
{\rm for\ } m+p>f,\cr
e_i^{(m)} e_{i\pm1}^{(p)} e_i^{(q)}&=\alpha^{(f-p)} e_i^{(m)} e_{i\pm1}^{(f-q)} \qquad{\rm if\ } m\geq q\cr
e_i^{(m)} e_{i\pm1}^{(p)} e_i^{(q)}&=\alpha^{(f-p)} e_{i\pm1}^{(f-m)} e_i^{(q)} \qquad{\rm if\ } m\leq q.\cr}$$ 
Again, the algebra depends only on one parameter, $\alpha$ and the number 
of points $n$. 
(Actually, the algebra is defined with $f-1$ parameters, and, we specialized all to be equal
so as to give a solution of the Yang--Baxter equation \r\Di.)
As in the two color case, the Hamiltonian is taken
to be,
$${\cal H}=\sum_{i=1}^n \sum_{a=1}^f  e_i^{(a)} k_a,\e$$
where the $k_a$ are some function of the color which depends only
on $\alpha$, and which guarantees that the model solves
the Yang--Baxter relation, ref. \r\Di. The Boltzman weights are 
all positive in the principal regime of the model which is defined by $\alpha>\sqrt2$.

Note that all of these algebras are nested inside each other.
Namely, the algebra for $r$ points is a subalgebra of the algebra
for $r+1$ points. This, very nontrivial fact, is going to be the basis 
for the subsequent discussion.

\section{The projectors}
Consider thus any of the algebras discussed above, $\cal A$. 
To be specific we can take $\cal A$ to be any of the Temperley--Lieb
or Fuss--Catalan algebras for some $f$, the number of colors, 
and some $n$, the number of points. Importantly, we do not include
the unit in this algebra, but just the generators $e_i^a$ obeying
eq. (5). 

Now, denote by $V_n$ a left and right unit of the algebra,
$$\manyeq{V_n e_i^{(a)}&=e_i^{(a)}\cr
  e_i^{(a)} V_n&=e_i^{(a)},\cr}$$
for all $i=1,2,\ldots,n$ and for all $a=1,2,\ldots,f$,
where we suppose that the unit indeed exists. Below we will give the 
explicit
expression for $V_n$ for the algebras we discuss.

Now, a standard result is that the unit, if it exists, is unique.
Suppose that $b$ is another unit (left or right; it does not matter).
Then, for a right unit, say,
$$b=b V_n=V_n,\e$$
which shows that the unit is always unique. From the unit,
we can form a projection operator, $Z_n$, by taking,
$$Z_n=1-V_n.\e$$
Since $V_n$ does not include a $1$, the projector $Z_n$ is nontrivial.
For the projector $Z_n$ we have,
$$e_i^{(a)} Z_n=Z_n e_i^{(a)}=0,\e$$
for all $i=1,2,\ldots,n$ and for all the $a=1,2,\ldots,f$.  

Before proceeding, let us give the explicit expression for the 
projectors $Z_n$ for the algebras on which we will concentrate here,
namely, the Temperley--Lieb and the Fuss--Catalan algebras.

For the TL algebra the projectors are given 
recursively by 
\REF\Temp{Temperley H N V, 1986 preprint, ``Potts models and 
related problems".}\r\Temp\ and reported by \REF\Mart{Martin P.P, J. Phys. A 21
(1988) 577.}\r\Mart,
$$Z_n=Z_{n-1} (1-a_n  e_n) Z_{n-1},\e$$
where
$$a_n={\sinh n\theta\over \sinh (n+1)\theta},\e$$
and $\theta$ obeys,
$$e^\theta+e^{-\theta}=\alpha,\e$$
and we define $Z_0=1$.
For this projector, it can be shown that eq. (10) is indeed
obeyed. The projectors are not defined for 
$$\alpha=2 \cos (\pi/N),\qquad {\rm for\ } N=2,3,\ldots,\e$$
where the $a_n$ are ill defined, for some $n$. These are exactly, the `minimal' models 
of this algebra, which are the only unitary models with $\alpha<2$,
and they correspond to conformal field theories. To avoid this problem,
we can take $\alpha$ to be close, but not identical, to any of
the minimal points. However, we shall not consider here these points in detail,
and in the sequel we shall assume that $\alpha$ is different from the values in eq. (14),
and that the projectors $Z_n$ are, thus, well defined. 

For the Fuss--Catalan algebra the expression for the projectors is
even simpler,
$$Z_n=\prod_{i=1}^n \left(1-{e_i^{(1)}\over \alpha}\right).\e$$
Here, it is entirely trivial that eq. (10) is obeyed, and that
$Z_n$ is the unique projector for the Fuss--Catalan algebra.
Here the projector is well defined for any $\alpha$, such that
$\alpha\neq0$, and we shall assume this to be the case.

Now, we wish to diagonalize the Hamiltonian of the system. I.e., 
to find solutions to the equation,
$${\cal H}\psi_n=E_n \psi_n,\e$$
where the $E_n$ are the energy levels and $\psi_n$ are the wave functions
(which we can take to be any of the operators in the algebra, acting
on some state, which is not of significance). What we will describe below is a method
of diagonalizing the Hamiltonian, by dividing the Hilbert space
into $n+1$ sectors which do not mix.

The discussion here is general for any algebra. So let
us assume that we have a descending chain of algebras ${\cal C}_r$,
such that each is a subalgebra of the other,
$${\cal C}_1\supset{\cal C}_2\supset \ldots\supset {\cal C}_n.\e$$
Assume also that for each algebra, we have, a unique projector,
$Z_r$, which is the projector of the ${\cal C}_r$ algebra.

Now, assume that we have the full algebra ${\cal C}_1$ for some  
number of points $n$. We can consider the chain of subalgebras
${\cal C}_r$ where $r=1,2,\ldots,n$. For concreteness
we take ${\cal C}_r$ to be the algebra of the points $e_i^{(a)}$,
where $i=r,r+1,\ldots,n$. Thus, we have also the chain of projectors,
$Z_r$ which obey, as above,
$$e_i^{(a)} Z_r=Z_r e_i^{(a)}=0,\e$$
for all $n\geq i\geq r$, and any $a$, and $Z_0=0$ by convention.

It follows, that we can define a left ideal, $I_r$, as the 
ideal of all the elements in ${\cal C}_1$, which
annihilate $Z_r$ from the left,
$$I_r=\left\{ \omega \bigg|\, \omega Z_r=0\right\}.\e$$
Evidently, $I_r$ is a left ideal. I.e., multiplying any element 
of $I_r$ from the left, by any element of the algebra, gives a result
which is still in the ideal,
$$\omega\in I_r \quad{\rm and \ } r\in {\cal C}_1\Longrightarrow r\omega\in I_r.\e$$
Because the algebras ${\cal C}_r$
are subalgebras of each other, it follows that the ideals also form
a chain,
$$I_0\supset I_1\supset I_2\supset \ldots \supset I_{n+1}=(0),\e$$
where $I_0$ is the entire algebra, $I_0={\cal C}_1$. 

It follows that we can define sectors of the algebra by taking
$$U_r=I_r\setminus I_{r+1},\e$$
that is, the elements of the $r$'th sector ($r=0,1,\ldots, n$),
are defined as all the elements obeying,
$$\omega Z_r=0,\qquad {\rm and\ } \omega Z_{r+1}\neq 0.\e$$
In this way we divide the algebra ${\cal C}_1$ into
$n+1$ distinct sectors, $U_r$, the union of which is the entire algebra, ${\cal C}_1$.

Now, our aim is to solve Schrodinger equation for some Hamiltonian,
which we denoted by $\cal H$, e.g., eq. (2) or eq. (4). Alternatively, we may solve
the equation in the algebra itself, without applying it to any 
state,
$${\cal H} \psi=E \psi,\e$$
where $\psi$ is any element of the algebra. We can assume now that 
$\psi$ is given by,
$$\psi=\sum C_s \beta_s,\e$$
where $\beta_s$ is in the $s$th ideal, $\beta_s\in I_s$, and $C_s$ are some numerical coefficients to be determined.
Since $I_s$ is a left ideal, acting by 
${\cal H}$ from the left leaves us within the ideal. So, we
can ignore the lower ideals, $I_m$ with $m<s$. Further, we can 
multiply from the right by the projector $Z_{r+1}$. It follows
that the Schrodinger equation becomes,
$${\cal H} \psi Z_{r+1}=E \psi Z_{r+1},\e$$
and the only non--trivial words that appear are the ones such that $\psi Z_r=0$ and
$\psi Z_{r+1}\neq 0$. It follows that we can diagonalize independently
only the words that are in the $r$th sector, $\psi_r\in U_r$, and in eq. (25) we
can assume that $C_s=0$ for $s\neq r$ where $r$ labels the sector. 

The prescription to do this is thus very simple. We take only words 
from the $r$'th sector. Then we act with $\cal H$ and eliminate
from the result any words which are in any higher sector, 
i.e., those in $I_{r+1}$.

This allows us to diagonalize the Hamiltonian sector by sector,
without any mixing with other sectors. Note, that this method
does not require integrability at all, and we can take any algebra
and any Hamiltonian. 

Consider then the first example of the Temperly--Lieb algebra.
It is not hard to see that a basis for the $r$th sector is
given by,
$$C^{m1,m2,\ldots, m_r}= K^{m_1}_1 K^{m_2}_2\ldots K^{m_r}_r,\e$$
where the word $K_{n}^m$ is defined as,
$$K_m^n=e_n e_{n-1}\ldots e_m,\e$$
and $n\geq m$. We impose also
$$0\leq m_1<m_2\ldots < m_r\leq n,\e$$
and $m_r\geq r$.
It is immediate that $C^{m_1,m_2,\ldots, m_r}$ lies in the ideal 
$I_r$. It is less immediate, but correct, that it forms a basis
for the $r$th sector, $U_r$, and that all the words of this type
are, in fact, independent of each other. 

For the Fuss--Catalan algebra the situation is more complicated,
because some of the $e_i^{(a)}$ commute with $e_{i\pm 1}^{(b)}$, for some
$a$ and $b$, eqs. (3,5).
Here we take for the basis elements, of the $r$th `prime sector', $P_r$
$$C^{m_1,m_2,\ldots, m_r}=K^{m_1}_1 K^{m_2}_2\ldots K^{m_r}_r,\e$$
where again the $m_i$ obey eq. (29) and the $K_n^m$ are defined by
$$K^n_m=e_n^{(a_n^m)} e_{n-1}^{(a_{n-1}^m)}\ldots e_m^{(a_m^m)},\e$$
where the $a_n^m$'s can take any value of the color,
$a_n^m=1,2,\ldots,f$. Here the sector $U_r$ is defined as the union
of all the prime sectors, $P_1,\ldots,P_r$ which obey,
$$\omega Z_r=0{\qquad\ } {\rm and\ } \omega Z_{r+1}\neq 0.\e$$
An additional complication is that now not all the words defined by
eq. (30) are linearly independent, and some words are needed to be 
discarded.  

In all the models, the $k=0$ sector contains just one element, the
unit element, $U_0=\{1\}$. Applying the Hamiltonian we see that,
$$H \psi=0,\e$$
up to elements that are in higher sectors. We conclude that there is
just one state in this sector, and its energy is zero, $E=0$.
This state can be the ground state, or not, depending on the algebra
and the range of its parameters, as will be discussed below.

Consider now the $k=1$ sector for the Temperley--Lieb model. We
take the Hamiltonian as in eq. (2). A basis for this sector is
given by the elements, eq. (27),
$$A_r=e_r e_{r-1}\ldots e_1,\e$$
where $r=1,2,\ldots,n$. We take, then, for the wave function,
$$\psi=\sum_{i=1}^n a_i A_i,\e$$
where the $a_i$ are some numerical coefficients to be determined.
Applying $H$ on $\psi$ and discarding any elements in higher
sectors, we arrive at the equation,
$$a_{i+1}+\alpha a_i+a_{i-1}=E a_i,\e$$
along with the boundary condition,
$a_0=a_{n+1}=0$. It is very easy to solve this equation exactly,
by taking the ansatz,
$$a_r=a q_1^r +b q_2^r,\e$$
where $a$ and $b$ are some coefficients determined by the boundary condition, and
$q_1$ and $q_2$ are the two solutions of the second order
equation,
$$q^2+(\alpha-E)q+1=0.\e$$ 
 
This equation (first derived by Levy, ref. \r\Levy) is in fact identical
to the Bethe Ansatz solution of the Heisenberg spin chain,
with one spin flipped, as derived by Alcaraz et al. 
\REF\Alcaraz{Alcaraz F.C., Barber M.N., Batchelor M.T., Baxter R.J., 
and Quispel G.R.W., J. Phys A 20 (1987) 6397.} (ref. \r\Alcaraz).
Showing that in this case, our method reproduces the Bethe Ansatz.

Actually, it is quite easy, for the Temperley--Lieb case,
to write the general solution for any sector $k$, using the
basis, eq. (27). Again, it can be seen to reproduce the results of
the Bethe Ansatz. Since the Bethe ansatz for this model
is fully known, we will omit the details. Actually, unlike
the Bethe ansatz, we do not require integrability and we can
solve the model for any Hamiltonian we wish, by exactly the same method.

\section{Numerical work.}

Let us turn now to the Fuss--Catalan model. We take here two colors, $f=2$,
and the Hamiltonian as in eq. (4).  
Again, for the $k=0$ sector, there is only the unit field and the energy
of it is $E=0$. For the $k=1$ sector, and for five points, $n=5$, we find $20$ independent elements, which are
all in the form eq. (30), and which obey,
$\omega Z_1=0$ and $\omega Z_2\neq0$, where $\omega$ is the operator (as otherwise they would be 
in an higher sector, $k>1$). We diagonalized the Hamiltonian for these $20$ elements, and
the results are listed in the table for $\alpha=1,1.5,1.8,2,2.5,3$. 

The results are as follows. For $\alpha>\sqrt2$ the ground state energy is $0$ and the next
state is at energy $E=1$. That is, in this regime, which is the principal regime,
we find that there is a mass gap of $E=1$ and the model is massive. Thus, we expect
the theory to be at the scaling limit, that of a massive integrable model.
It is possible to check also the $k=2$ sector. The results persist also in this sector,
and we find only states with the energy $E\geq 1$. Note that all the eigenvalues in this regime are positive.
We do not enclose the calculations for the sector $k=2$ for the sake of brevity. We also run the program 
on other number of points, up to $n=11$, and the results above, indeed, persist in these
calculations.

For $0<\alpha<\sqrt2$ regime the situation is different. Here, for the $k=1$ sector, we always find complex eigenvalues,
indicating that the model is not unitary. There is one important exception to this which is for $\alpha=1$.
For this point we get for $k=1$ only non--negative integers (see the table). However, for $k=2$ we find
some (non--integral) negative eigenvalues. 
For example, for $n=4$ points, in the $k=2$ sector, the lowest energy state, i.e., the ground state of this 
sector, we find,
$$E={1-\sqrt5\over 2}=-0.618034...,\e$$
and it is the only negative energy state in this sector.
We conclude that for $\alpha=1$ the ground state energy is negative,
and that there is probably no mass gap. This theory is analogous to the $\alpha<2$ unitary minimal model
for the TL algebra (where $\alpha$ is given by eq. (14)). We conjecture then that the $\alpha=1$ model
is some conformal field theory in the scaling limit. However, more work is needed to determine
precisely the RCFT. 

We expect these results to persist for more than two colors. Again in the principal regime
$\alpha>\sqrt2$ we expect to have a zero energy ground state and a mass gap of $1$.
Again, $\alpha=1$ is the only unitary point for $\alpha<\sqrt2$, and it is conjectured
to be some conformal field theory in the scaling limit. We thus have a series
of RCFT's or $\alpha=1$ and some number of colors, $f$, where $f=2,3,4,\ldots$.
It is left to future work to determine exactly the RCFT's. 

\section{Conclusions.}
We described here an algebraic method for diagonalizing the Hilbert
space of solvable lattice models. We exemplified the method on the Temperley--Lieb and
Fuss--Catalan models. We expect this method to be of interest
for other models obeying such a chain of algebras. We hope that
this will be beneficial to the study of lattice models in two 
dimensions along with their interplay with quantum field theory.

\overfullrule=0pt

%
%
\setbox\strutbox=\hbox{\vrule height12pt depth8pt width0pt}
\midinsert
\line{\hfill Table.\hfill}
\line{\hfill The spectrum for $n=5$ and $k=1$.\hfill}
\vskip15pt
\line{\hfill
\vbox{\offinterlineskip\hrule
\halign{&\vrule#&
  \strut\quad\hfil$#$\hfil\quad\cr
height2pt&\omit&&\omit&&\omit&&\omit&&\omit&&\omit&\cr
&\alpha=1&&\alpha=1.5&&\alpha=1.8&&\alpha=2&&\alpha=2.5&&\alpha=3&\cr
\noalign{\hrule}
height2pt&\omit&&\omit&&\omit&&\omit&&\omit&&\omit&\cr
&2&& 17.8301&& 6.4410&& 5.3364 && 
  4.4705 && 4.2857&\cr
&2 && 17.4643 && 
  5.9635 && 5. && 4.3268 && 3.9383&\cr
&2 && 15.6332 && 5.7745 && 5. &&  4.1244 && 
  3.7901&\cr
&2&& 14.7850 && 5.6129 && 
  4.5647 && 3.7744 && 3.5469&\cr 
& 2 &&  12. &&  5.0815 && 4.3660 && 3.4705 && 
  3.2857&\cr
&2 && 11.4721 && 5.0134 && 
  4.1037 && 3.4577 && 3.1428&\cr
&2 && 11. && 4.6129 && 4. && 3.2352 && 3.0943&\cr
& 1 && 10.6190 && 4.1216 && 3.5000 && 
  3.2049 && 3.0312&\cr
&1 && 7.0000 && 3.8064 && 3.4595 && 3.1544 && 
  2.8565&\cr
&1 && 6.4306 && 3.5165  
 && 3.3240 && 2.8456 && 2.7387&\cr 
 & 1 && 5.4417 && 3.1236 && 2.7772 && 
  2.6961 &&  2.6384&\cr
&1 &&  3.2737 && 
  2.5994 && 2.6339 && 2.5814 && 
  2.6050&\cr
&1 && 2.5278 &&  2.4557 && 
  2.5000 &&  2.4598 && 2.5387&\cr 
&1 && 2.3667 && 2.2941 && 2.3395 && 
  2.4520 &&  2.3233&\cr
&1 && 2.0939 && 
  2.2682 && 2.1876 && 2.0910 && 
  2.0559&\cr
&1 && 1.6871 && 1.7383 && 
  1.7623 && 1.8047 && 1.8332&\cr
&1 && 1.2134 && 1.4311 && 1.5000 && 
  1.60438 && 1.6678&\cr
&1 && 1.0972 && 
  1.2822 && 1.3567 && 1.4794 && 
  1.5582&\cr
&1 && 1.0636 && 1.2171 && 
  1.2881 && 1.4129 && 1.4971&\cr 
&0 && 1. && 1.&& 1. && 1. && 1.&\cr
height2pt&\omit&&\omit&&\omit&&\omit&&\omit&&\omit&\cr
}\hrule}
\hfill}
\setbox\strutbox=\hbox{\vrule height8.5pt depth3.5pt width 0pt}
\vskip20pt\endinsert
\ack
We thank the Einstein Center for support.
\refout
\bye